\documentclass[twocolumn,showpacs,preprintnumbers,amsmath,amssymb]{revtex4}
\usepackage{dcolumn}
\usepackage{bm}
\usepackage{amsmath}
\usepackage{color}

\def\cor#1{ #1 }     

\def\ket#1{ $ \left\vert  #1   \right\rangle $ }
\def\ketm#1{  \left\vert  #1   \right\rangle   }
\def\bram#1{  \left\langle  #1   \right\vert   }
\def\spr#1#2{  \left\langle #1 \left\vert \right. #2 \right\rangle   }
\def\etal{\textit{et al.}}

\begin{document}

\title{Universal quantum Controlled-NOT gate}

\author{Michael Siomau}
 \email{siomau@physi.uni-heidelberg.de}
 \affiliation{Max-Planck-Institut f\"{u}r Kernphysik, Postfach
103980, D-69029 Heidelberg, Germany}
 \affiliation{Physikalisches
Institut, Heidelberg Universit\"{a}t, D-69120 Heidelberg, Germany}

\author{Stephan Fritzsche}
 \affiliation{Department of Physics, P.O.~Box 3000, Fin-90014
University of Oulu}
 \affiliation{GSI Helmholtzzentrum f\"{u}r Schwerionenforschung, D-64291
Darmstadt, Germany}

\date{\today}

\begin{abstract}
\cor{An investigation of} an optimal universal unitary
\textsl{Controlled-NOT} gate that performs a specific
\cor{operation} on two unknown \cor{states of} qubits taken from a
great circle of the Bloch sphere is presented. The deep analogy
between the optimal universal \textsl{C-NOT} gate and the
`equatorial' quantum cloning machine (QCM) is shown. In addition,
possible applications of the universal \textsl{C-NOT} gate are
briefly discussed.
\end{abstract}


\maketitle

\section{\label{sec:1} Introduction}

Manipulation of information encoded in states of quantum systems has
remarkable advantages compared to classical information processing
\cite{Nielsen:00}. Unlike classical information, however, there is a
fundamental limitation on the basic operations that one can perform
on quantum systems. This limitation is known as the
\textit{non-cloning} theorem \cite{Wootters:82} and has \cor{a
manifest impossibility of conducting} several \textit{exact}
operations in quantum information theory, such as `cloning'
\cite{Wootters:82}, `inversion' \cite{Gisin:99} and `entangling'
\cite{Buzek:00} of unknown quantum states. If one does not demand
these operations to be perfect, it is possible to construct quantum
devices that provide the required operations approximately. Many
examples of approximate operations on \cor{states of} quantum
systems have been shown in the last decades: universal symmetric
\cite{Buzek:96} and asymmetric \cite{Cerf:98} quantum cloning
machines (QCM's), universal \textsl{NOT} gate \cite{Gisin:99} as
well as universal symmetric \cite{Buzek:00} and asymmetric
\cite{Horoshko:04} entanglers.

In this work we pay attention to a quantum operation,
\textsl{Controlled-NOT} gate, that plays an important role in
quantum information theory and especially in quantum computing
\cite{Nielsen:00}. Although this gate is usually associated with a
computational basis, for a deeper understanding of the fundamental
principles of operating with quantum systems, it is essential to
investigate a universal operation that is basis independent. While
the impossibility of constructing an \textit{exact} \cor{basis
independent} \textsl{C-NOT} gate has already been shown
\cite{Pati:02}, we shall present a universal \textsl{C-NOT} gate
that provides an approximate transformation on two (input) qubits in
unknown quantum states. At first, however, we shall discuss the
universal \textsl{NOT} gate that is an essential part of universal
\textsl{C-NOT} gate. \cor{Although} an exact universal \textsl{NOT}
gate for a qubit in an unknown input state does not exist
\cite{Gisin:99}, we show that there is an exact universal
\textsl{NOT} gate for an unknown qubit state chosen from a great
circle of the Bloch sphere. With the help of this exact \textsl{NOT}
gate, we then construct \cor{an} optimal universal \textsl{C-NOT}
gate for two \cor{unknown states of qubits} chosen from the great
circle of the Bloch sphere. \cor{This optimal \textsl{C-NOT} gate
has similar structure to the `equatorial' QCM \cite{Bruss:00} while
the fidelity between the ideal and the actual output states of the
universal \textsl{C-NOT} gate equals $F = 1/2 + \sqrt{1/8}$ for both
qubits.}

This paper is organized as follows. In the next section we \cor{
discuss universal \textsl{NOT} operation \cite{Gisin:99} and
construct an exact universal \textsl{NOT} gate for an unknown qubit
state taken from a great circle of the Bloch sphere. Having the
exact \textsl{NOT} gate,} we then suggest an `idealized' universal
\textsl{C-NOT} operation which, however, can not be achieved due to
the non-cloning principle. In Section~\ref{sec:3} we present an
optimal (approximate) universal \textsl{C-NOT} gate for unknown
input states of qubits taken from a great circle of the Bloch
sphere. Finally, several concluding remarks are drawn in
Section~\ref{sec:4}.

\section{\label{sec:2} Universal quantum gates}

At the beginning of discussion of the universal quantum gates we
note that we always use the Bloch sphere representation of a qubit
state. A pure state of a qubit can be written as
$\ketm{\psi} \,=\, \cos{\frac{\theta}{2}} \ketm{0} +
\sin{\frac{\theta}{2}} e^{i\varphi}\ketm{1}$,
and with $\ketm{0}$ and $\ketm{1}$ being computational basis states.
In this representation, the parameters $\theta$ and $\varphi$ take
values in the range $0 \,\le\, \theta \,\le\, \pi$ and $0 \,\le\,
\varphi \,<\, 2\pi$, respectively, and we shall often use this Bloch
sphere in order to visualize the states of interest.

\subsection{\label{subsec:2.1} NOT gate}

Let us start our discussion of the universal quantum gates by
recalling the properties of the universal quantum \textsl{NOT}
\cor{operation} \cite{Gisin:99}. According to the definition, for a
given input qubit state \ket{\psi}, this gate generates the
orthogonal state \ket{\psi^\bot} at the output, i.e.
\begin{equation}
\label{not}
  \textsl{NOT} \ketm{\psi} = \ketm{\psi^\bot} \, ,
\end{equation}
so that $\spr{\psi} {\psi^\bot} \equiv 0$. An exact unitary
transformation (\ref{not}) for an arbitrary input qubit state
\ket{\psi} does not exist \cite{Gisin:99}. To provide this
transformation approximately, Bu\v{z}ek \etal \, considered an
ensemble of $N$ input qubits that are prepared in the state
\ket{\psi}. It was shown that the transformation (\ref{not}) can be
performed approximately on the ensemble with fidelity $F =
\spr{\psi^\bot}{\rho \vert \psi^\bot} = (N + 1)/(N + 2)$ between the
approximate output $\rho$ of the transformation and the ideal output
\ket{\psi^\bot}. If the input ensemble consists of a single state
\ket{\psi}, the universal unitary \textsl{NOT} transformation has
the structure \cite{Gisin:99}
\begin{eqnarray}
\label{U-not}
   \ketm{\psi} \ketm{X} \longrightarrow
\sqrt{\frac{2}{3}} \ketm{\psi^\bot} \ketm{A} \;+\;
\sqrt{\frac{1}{3}} \ketm{\psi} \ketm{B} \, ,
\end{eqnarray}
where \ket{X}, \ket{A} and \ket{B} are the state vectors of the
device (with an auxilliary system). The fidelity between the
approximate output of the transformation (\ref{U-not}) and the ideal
output \ket{\psi^\bot} equals $F_{NOT} = 2/3$.

The universal \textsl{NOT} transformation (\ref{U-not}) for an
arbitrary input qubit state has low fidelity. To improve the
fidelity of the universal \textsl{NOT} transformation one may
consider a restricted set of input states, for example, a
one-dimensional subspace of the two-dimensional Hilbert state space
of a qubit. Using the Bloch sphere representation of the qubit
state, a one-dimensional subspace can be visualized as an
intersection of the Bloch sphere with a plane. Let us consider the
one-dimensional subspace, \textit{the main circle}, that is formed
by the intersection of the Bloch sphere with \textit{x-z} plane. An
arbitrary state of a qubit in this circle can be parameterized as
\begin{eqnarray}
\label{states}
   \ketm{\psi} & = & \cos{\frac{\theta}{2}} \ketm{0}
                \pm \sin{\frac{\theta}{2}}\ketm{1} \, .
\end{eqnarray}

We found that for an arbitrary input state (\ref{states}) the
operator
\begin{equation}
\label{main_cir-NOT}
    \textsl{NOT} = - i \sigma_y =
    \begin{pmatrix}
          0 & -1 \\
          1 & 0
    \end{pmatrix} \,
\end{equation}
provides an \textit{exact} \textsl{NOT} transformation, i.e.
$\spr{\psi} {\textsl{NOT} |\, \psi} \equiv 0$. This operator
introduces an exact universal \textsl{NOT} gate for the input state
(\ref{states}). Moreover, from the symmetry of the Bloch sphere it
follows that a universal \textsl{NOT} gate can be constructed for
any restricted one-dimensional set of input states.

Knowing the \cor{expressions for} universal \textsl{NOT}
\cor{transformations} for an arbitrary input state (\ref{U-not}) and
for an input state from the main circle (\ref{main_cir-NOT}), one
may use one of these transformations in order to construct a
universal \textsl{C-NOT} gate. Since the universal gate
(\ref{U-not}) has low fidelity $F_{NOT} = 2/3 \approx 0.67$ we shall
not discuss a universal \textsl{C-NOT} gate based on this
approximate transformation \cite{comment}. Instead, we shall focus
on the universal \textsl{C-NOT} gate that is based on an exact
universal \textsl{NOT} gate (\ref{main_cir-NOT}). Of course, such a
\textsl{C-NOT} gate \cor{is restricted by the input states
(\ref{states}) of qubits taken from the main circle of the Bloch
sphere.}

\subsection{\label{subsec:2.2} C-NOT gate}

A quantum \textsl{Controlled-NOT} gate provides a unitary
transformation on two qubits, one of which is called
\textit{control} and the other -- \textit{target}. According to the
definition \cite{Nielsen:00}, the gate leaves the meaning of the
target qubit unchanged, if the control qubit is given in the state
\ket{0}. If the control qubit is in the state \ket{1}, the gate
performs a \textsl{NOT} operation on the target qubit. If the
control qubit is given in a superposed state, the quantum
\textsl{C-NOT} gate can be defined on a computational basis as
\cite{Nielsen:00}
\begin{equation}
\label{standart-CNOT}
 U = \ketm{0}\bram{0}_c\otimes I_t + \ketm{1}\bram{1}_c \otimes
 (\sigma_x)_t \, ,
\end{equation}
where $\sigma_x = \ketm{0}\bram{1} + \ketm{1}\bram{0}$. The gate
(\ref{standart-CNOT}) leaves the states of the control and the
target qubit separable, if the control qubit is in one of the basis
states \ket{0} or \ket{1}, and creates entanglement between the
control and the target qubits otherwise.

In contrast to the definition (\ref{standart-CNOT}), the universal
(basis independent) \textsl{C-NOT} gate should always leave the
states of the control and the target qubits separable. Indeed, a
superposed state of the control qubit in a given basis can be always
transformed in one of the basis states into a new basis by means of
a basis transformation. In this new basis the states of the control
and the target qubits are separable according to the definition
\cor{given} above. \cor{On the other hand,} the basis independent
\textsl{C-NOT} gate should be invariant with regard to a basis
transformation. \cor{Therefore}, for a given superposed input state
of the control qubit, the output states of the control and the
target qubits are separable.

As mentioned in the previous section, we are going to construct a
universal \textsl{C-NOT} transformation for the states of the
control and the target qubits taken from the main circle of the
Bloch sphere. Let us introduce the following notations for these
qubits
\begin{eqnarray}
\label{control}
   \ketm{\psi_\pm}_c & = & \cos{\frac{\theta}{2}} \ketm{0}_c
                \pm \sin{\frac{\theta}{2}} \ketm{1}_c \, ,
\\[0.1cm]
\label{target}
   \ketm{\chi_\pm}_t & = & \cos{\frac{\phi}{2}} \ketm{0}_t
                \pm \sin{\frac{\phi}{2}} \ketm{1}_t \, ,
\end{eqnarray}
where $\ketm{\psi_\pm}_c$ and $\ketm{\chi_\pm}_t$ denote the states
of the control and the target qubits respectively. Although the
state $\ketm{\psi_\pm}_c$ of the control qubit is given in a
superposition of the two basis states $\ketm{0}_c$ and $\ketm{1}_c$,
it is of course sufficient to know the \cor{\textsl{C-NOT}}
transformation of just the basis in order to obtain a proper
transformation for states \cor{(\ref{control})-(\ref{target})}. For
the input states $\ketm{0}_c$ and $\ketm{1}_c$ of the control qubit,
the universal unitary \textsl{C-NOT} should perform the
transformation
\begin{eqnarray}
\label{CNOT-1}
   \ketm{0}_c \ketm{\chi_\pm}_t \ketm{Q}_d & \longrightarrow &
   \ketm{0}_c \ketm{\chi_\pm}_t \ketm{Q_0}_d \, ,
\\[0.1cm]
\label{CNOT-2}
   \ketm{1}_c \ketm{\chi_\pm}_t \ketm{Q}_d & \longrightarrow &
   \ketm{1}_c \ketm{\chi_\pm^\bot}_t \ketm{Q_1}_d \, ,
\end{eqnarray}
\cor{as required by the definition}. The state vectors $\ketm{Q}_d$,
$\ketm{Q_0}_d$ and $\ketm{Q_1}_d$ denote the initial and the final
states of the device that provides this transformation. The output
state $\ketm{\chi_\pm^\bot}_t$ of the target qubit is orthogonal to
the input target qubit state $\ketm{\chi_\pm}_t$ and is obtained by
applying the \textsl{NOT} gate (\ref{main_cir-NOT}) to the input
state (\ref{target}), i.e. $\ketm{\chi_\pm^\bot}_t = \textsl{NOT}
\ketm{\chi_\pm}_t$. If the state of the control qubit is given in
the superposed state (\ref{control}), the universal unitary
\textsl{C-NOT} transformation should leave the states of the control
and the target qubits separable \cor{while} performing some
transformation $f(\psi, \chi)$ on the target qubit, i.e.
\begin{eqnarray}
\label{ideal-CNOT}
   \ketm{\psi}_c \ketm{\chi}_t \ketm{Q}_d & \longrightarrow &
   \ketm{\psi}_c \ketm{f(\psi, \chi)}_t \ketm{Q_\psi}_d \, ,
\end{eqnarray}
where the function $f(\psi, \chi)$ is related to the original state
$\ketm{\chi}_t$ by a unitary transformation $\ketm{f(\psi, \chi)}_t
= U(\psi)\:\ketm{\chi}_t$. On the other hand, making a superposition
of Eqns.~(\ref{CNOT-1})-(\ref{CNOT-2}) we obtain
\begin{eqnarray}
\label{CNOT-superposition}
     & & \left( \cos{\frac{\theta}{2}} \ketm{0}_c
     + \sin{\frac{\theta}{2}} \ketm{1}_c \right) \ketm{\chi_\pm}_t \ketm{Q}_d
     \longrightarrow
\\[0.1cm]
     & & \hspace*{0cm}
       \cos{\frac{\theta}{2}} \ketm{0}_c \ketm{\chi_\pm}_t
       \ketm{Q_0}_d \:+\: \sin{\frac{\theta}{2}} \ketm{1}_c
       \ketm{\chi_\pm^\bot}_t \ketm{Q_1}_d  .
\nonumber
\end{eqnarray}

Let us analyze this transformation (\ref{CNOT-superposition}) in
order to specify the function $f(\psi, \chi)$ in the transformation
(\ref{ideal-CNOT}). Suppose one has two qubits prepared in the
states $\ketm{\psi_0}_c = \cos{\frac{\theta_0}{2}} \ketm{0}_c +
\sin{\frac{\theta_0}{2}} \ketm{1}_c$ and $\ketm{\chi_0}_t$
respectively. If one performs the transformation
(\ref{CNOT-superposition}) on them, so that the qubit
$\ketm{\psi_0}_c$ is the control and the qubit $\ketm{\chi_0}_t$ --
the target, the two-qubit state
\begin{eqnarray}
       \cos{\frac{\theta_0}{2}} \ketm{0}_c \ketm{\chi_0}_t  \:+\:
       \sin{\frac{\theta_0}{2}} \ketm{1}_c \ketm{\chi_0^\bot}_t \, ,
\end{eqnarray}
is obtained at the output, as it follows from
Eqns.~(\ref{CNOT-1})-(\ref{CNOT-2}) and (\ref{CNOT-superposition}).
Making a projective measurement on the target qubit in the
$\{\ketm{\chi_0}_t, \ketm{\chi_0^\bot}_t\}$ basis one obtains the
outcomes $\ketm{\chi_0}_t$ and $\ketm{\chi_0^\bot}_t$ with
probabilities $\cos^2{\frac{\theta_0}{2}}$ and
$\sin^2{\frac{\theta_0}{2}}$ \cor{respectively}. \cor{For the
universal \textsl{C-NOT} gate (\ref{ideal-CNOT}), we suggest this
transformation take the following structure}
\begin{eqnarray}
\label{pretrans}
      & &   \ketm{\psi_+}_c \ketm{\chi_\pm}_t  \ketm{Q}_d
      \longrightarrow
\\[0.1cm] & & \hspace*{0cm} \ketm{\psi_+}_c \;
      \left( \cos{\frac{\theta}{2}} \ketm{\chi_\pm}_t +
      \sin{\frac{\theta}{2}} \ketm{\chi_\pm^\bot}_t \right)
      \ketm{Q_\psi}_d \, .
\nonumber
\end{eqnarray}
On the right hand side of this transformation (\ref{pretrans}), the
control qubit is left without changes as is required by
Eqn.~(\ref{ideal-CNOT}) while the unitary transformation
\begin{equation}
    U(\psi) =
    \begin{pmatrix}
          \, \cos{\frac{\theta}{2}} & -\sin{\frac{\theta}{2}} \\
          \sin{\frac{\theta}{2}} & \cos{\frac{\theta}{2}}
    \end{pmatrix} \,
\end{equation}
is to be performed on the target qubit $\ketm{\chi}_t$. After simple
algebraic manipulation we find that the universal \textsl{C-NOT}
operation can be written as
\begin{eqnarray}
\label{CNOT}
   & &   \ketm{\psi_+}_c  (\cos{\frac{\phi}{2}} \ketm{0}_t
         \pm \sin{\frac{\phi}{2}} \ketm{1}_t)  \ketm{Q}_d \longrightarrow
\\[0.1cm] & & \hspace*{0cm}
         \ketm{\psi_+}_c (\cos{\frac{\phi - \theta}{2}} \ketm{0}_t
         \pm \sin{\frac{\phi - \theta}{2}} \ketm{1}_t)
         \ketm{Q_\psi}_d \, ,
\nonumber
\end{eqnarray}
where we have shown the states of the target qubit before and after
the transformation explicitly. The transformation (\ref{CNOT})
leaves the control qubit without changes and rotates the target
qubit on the angle $\theta$ clockwise. We note that if the state of
the control qubit is given in the state $\ketm{\psi_-}_c$, the
transformation (\ref{CNOT}) rotates the target qubit
counterclockwise to the angle $\theta$. It is also remarkable that
the state of the output target qubit depends only on the difference
$\phi - \theta$ and does not depend on a particular basis (as it
should be for a basis independent transformation).

The transformation (\ref{CNOT}) introduces the `idealized' universal
\textsl{C-NOT} gate that can not be performed exactly due to the
non-cloning principle \cite{Pati:02}. Indeed, to perform the
rotation of the target qubit, the device needs to obtain some
information about the input state $\ketm{\psi}_c$ of the control
qubit. The non-cloning principle implies that any information can
not be obtained from the unknown state $\ketm{\psi}_c$ without
changing the state. Nevertheless, in the next section we shall
construct an optimal universal \textsl{C-NOT} gate that provides the
transformation (\ref{CNOT}) approximately with constant fidelity for
both the control and the target qubit states taken from the main
circle.

\section{\label{sec:3} Explicit form of the universal C-NOT gate}

To obtain an explicit form of the (approximate) universal
\textsl{C-NOT} transformation (\ref{CNOT}), let us consider the most
general quantum transformation for two-qubit (control + target)
state which can be cast in the form
\begin{eqnarray}
\label{general}
   \ketm{0}_c \ketm{\chi_\pm}_t \ketm{Q}_d  & \longrightarrow &
   \sum_{m,n=0}^1 \, \ketm{m}_c \ketm{n}_t \ketm{Q_{mn}}_d \, ,
\nonumber \\[0.1cm]
   \ketm{1}_c \ketm{\chi_\pm}_t \ketm{Q}_d  & \longrightarrow &
   \sum_{m,n=0}^1 \, \ketm{m}_c \ketm{n}_t \ketm{Q_{mn}^\prime}_d \, ,
\end{eqnarray}
where $\ketm{Q}_d$ denotes again the initial state of the device.
Once, the transformation has been performed, $\ketm{m}_c$ and
$\ketm{n}_t$ denote the output basis states of the control and
target qubits, while $\ketm{Q_{mn}}_d$ and $\ketm{Q_{mn}^\prime}_d$
are the corresponding states of the apparatus. In order to ensure
that the transformation (\ref{general}) is \textit{unitary},
\begin{eqnarray}
\label{unitary}
   \sum_i \, c_i \ketm{i}_{ct} \ketm{Q}_d &\longrightarrow&
   \sum_{i,\lambda} \, c_i U_{i\lambda}\ketm{\lambda}_{ctd} \, ,
\end{eqnarray}
for all possible input states, i.e.\ for $\ketm{i} \,=\,
\{\ketm{0}_c\ketm{\chi_\pm}_t,$ $ \ketm{1}_c\ketm{\chi_\pm}_t \}$,
the three-partite basis $\{ \ketm{\lambda}_{ctd} \}$ refers to a
complete and orthonormal basis for the overall system (qubits c, \,
t + device). Thus, the requested unitarity $UU^\dag \,=\, 1$ of the
transformation (\ref{general}) implies the conditions
\begin{eqnarray}
\label{general-cond}
   \sum_{m,n=0}^1  \, {}_d \spr{Q_{mn}}{Q_{mn}}_d & = &
   \sum_{m,n=0}^1  \, {}_d \spr{Q_{mn}^\prime}{Q_{mn}^\prime}_d
   \; = \; 1 \, ,
   \nonumber  \\
   \sum_{m,n=0}^1  \, {}_d \spr{Q_{mn}}{Q_{mn}^\prime}_d & = & 0 \, .
\end{eqnarray}

For any explicit construction of transformation (\ref{general}), we
must therefore `determine' the final states $\ketm{Q_{mn}}_d$ and
$\ketm{Q_{mn}^\prime}_d$ of the device in line with the conditions
(\ref{general-cond}) and an additional optimality condition that
specify a particular transformation (\ref{general}). Let us require
that the unitary transformation (\ref{general}) realizes the
universal \textsl{C-NOT} gate (\ref{CNOT}) with maximal average
fidelity between the input and the output states of the control as
well as the target qubits. The average fidelity is defined as an
integral of a fidelity function over a set of states and is given by
\cite{Audenaert:02,Fiurasek:03}
\begin{equation}
 \label{average fidelity}
\overline{F} = \int_\Omega \: \frac{d \phi}{A} \; F_c(\phi) =
\int_\Omega \: \frac{d \theta}{A} \; F_t(\theta) \, ,
\end{equation}
where $A$ is a normalization factor. The fidelity functions
$F_c(\phi)$ and $F_t(\theta)$ are defined as $F_c(\phi) = {}_c
\spr{\psi^{id}}{\rho^{out}_c| \psi^{id}}_c$ and $F_t(\theta) = {}_t
\spr{\chi^{id}}{\rho^{out}_t| \chi^{id}}_t$, where \cor{
$\ketm{\psi^{id}}_c$ and $\ketm{\chi^{id}}_t$ denote the ideal
output states of the control and the target qubits while}
$\rho^{out}_c$ and $\rho^{out}_t$ are the actual \cor{(approximate)}
output states of the control and the target qubits from the
transformation (\ref{general}) respectively. In the expression
(\ref{average fidelity}) the integration of the fidelity functions
is to be done over all the states $\Omega$ from the main circle of
the Bloch sphere.

To find the maximum of the average fidelity (\ref{average fidelity})
we used the general \textit{method of semidefinite programming}
\cite{Audenaert:02,Fiurasek:03} which allows one to find the optimal
unitary transformation with regard to some specific conditions.
Using this method we found the optimal universal unitary
\textsl{C-NOT} gate for the input states of the control and the
target qubits taken from the main circle of the Bloch sphere. This
gate \cor{can be given in a chosen basis} by the transformation
\begin{small}
\begin{eqnarray}
\label{CNOT-optimal-1}
   \ketm{0}_c \ketm{\chi_\pm}_t \ketm{Q}_d
   &\longrightarrow & \left( \frac{1}{2} + \sqrt{\frac{1}{8}} \right)
   \ketm{0}_c \ketm{\chi_\pm}_t \ketm{0}_d
\nonumber\\[0.1cm]
   & & \hspace*{-2cm} \:+\: \sqrt{\frac{1}{8}} \left( \ketm{0}_c
   \ketm{\chi_\pm^\bot}_t + \ketm{1}_c \ketm{\chi_\pm}_t \right) \,
   \ketm{1}_d
\nonumber\\[0.1cm]
   & & \hspace*{-2cm} \:+\: \left( \frac{1}{2} - \sqrt{\frac{1}{8}} \right)
   \ketm{1}_c \ketm{\chi_\pm^\bot}_t \ketm{0}_d \, ,
\\[0.2cm]
\label{CNOT-optimal-2}
   \ketm{1}_c \ketm{\chi_\pm}_t \ketm{Q}_d
   &\longrightarrow & \left( \frac{1}{2} + \sqrt{\frac{1}{8}} \right)
   \ketm{1}_c \ketm{\chi_\pm^\bot}_t \ketm{1}_d
\nonumber\\[0.1cm]
   & & \hspace*{-2cm} \:+\: \sqrt{\frac{1}{8}} \left( \ketm{0}_c
   \ketm{\chi_\pm^\bot}_t + \ketm{1}_c \ketm{\chi_\pm}_t \right) \,
   \ketm{0}_d
\nonumber\\[0.1cm]
   & & \hspace*{-2cm} \:+\: \left( \frac{1}{2} - \sqrt{\frac{1}{8}} \right)
   \ketm{0}_c \ketm{\chi_\pm}_t \ketm{1}_d \, .
\end{eqnarray}
\end{small}
\cor{This transformation is invariant with regard to a basis
transformation by construction.} For the transformation
(\ref{CNOT-optimal-1})-(\ref{CNOT-optimal-2}) the fidelity between
the ideal output and the actual output for the states of the control
as well as the target qubits equals $F = 1/2 + \sqrt{1/8}$ and is
constant for arbitrary input states of the control and target qubits
taken from the main circle of the Bloch sphere.

The transformation (\ref{CNOT-optimal-1})-(\ref{CNOT-optimal-2}) has
similar structure to the `equatorial' QCM \cite{Bruss:00}. This
similarity has an important implication. The `idealized' universal
\textsl{CNOT} transformation (\ref{CNOT}) can be formally treated as
a two-step transformation. The first stage of the device provides
the cloning transformation on the input control qubit, the second
stage rotates the state vector of the copy in the main circle over
the angle $\phi$ which describes the state of the target qubit.
\cor{While the first stage (cloning) transformation is strongly
restricted by the non-cloning principle, there are no limitations on
the second stage transformation.} Thereby the problem to find an
optimal \textsl{C-NOT} transformation for the input states of the
qubits taken from the main circle reduces to a search for the
optimal cloning transformation for such input states. Since the
`equatorial' QCM is the optimal cloning transformation for the input
states from the main circle \cite{Bruss:00}, it is not surprising
that the universal \textsl{C-NOT} transformation
(\ref{CNOT-optimal-1})-(\ref{CNOT-optimal-2}) has a structure
similar to the `equatorial' QCM.

\section{\label{sec:4} Concluding remarks}

Unlike the well-known basis dependent \textsl{C-NOT} gate
(\ref{standart-CNOT}), we have presented an optimal universal
\textsl{C-NOT} gate that performs the transformation
(\ref{ideal-CNOT}) \cor{approximately} on two unknown input qubits
taken form the main circle of the Bloch sphere. The obtained
universal \textsl{C-NOT} gate provides the transformation
(\ref{CNOT-optimal-1})-(\ref{CNOT-optimal-2}) on the input states of
the qubits with constant fidelity $F = 1/2 + \sqrt{1/8}$ between the
ideal output and the actual output for both the control and the
target qubits. Moreover, we have shown the analogy between universal
\textsl{C-NOT} gate and QCM which makes possible the construction of
\cor{\textsl{C-NOT} gates with various properties}. For example, one
may construct a universal asymmetric \textsl{C-NOT} gate that
provides the transformation (\ref{ideal-CNOT}) with different
fidelities for the control and the target qubits $F_c \neq F_t$.
This universal `asymmetric' \textsl{C-NOT} gate represents an analog
of the universal asymmetric QCM \cite{Cerf:98}. Another possibility
is to construct a universal \textsl{C-NOT} gate for the input states
(of control and target qubits) from a small circle on the Bloch
sphere that is formed by a plane that crosses the sphere away from
its center (similar QCM was considered by Fiur\'{a}\v{s}ek
\cite{Fiurasek:03}). Finally, one may consider the possibility to
construct a universal probabilistic \textsl{C-NOT} gate that allows
one to perform the transformation (\ref{ideal-CNOT}) exactly with a
distinct probability \cite{Duan:98,Pati:99}.

Besides the pure theoretical interest, the universal \textsl{C-NOT}
gate may find its applications in quantum communication and quantum
computing, since it has some advantages compared to the basis
dependent \textsl{C-NOT} gate (\ref{standart-CNOT}). Apart from the
fact that the universal \textsl{C-NOT} gate operates with unknown
input \cor{states of} qubits, it may efficiently operate with mixed
input states since \cor{the optimal cloning transformation for mixed
input states has already been developed} \cite{Scarani:05}. It has
recently been shown that quantum computing with mixed quantum states
has advantages over the best possible classical computation
\cite{Jozsa:03} and may, in particular, provide the computational
speed up of Deutsch-Jozsa and Simon problems in comparison to the
best known classical algorithms \cite{Biham:04}. The universal
\textsl{C-NOT} gate introduces a basic element for possible schemes
of quantum computation \cor{based on} mixed states.

We also hope that the universal \textsl{C-NOT} gate can be realized
experimentally with good accuracy, since efficient experimental
realizations of QCM have already been demonstrated
\cite{Scarani:05}. For example, an optical implementation of the
universal QCM \cite{Buzek:96} based on parametric down-conversion
has been achieved with fidelity $0.810 \pm 0.008$
\cite{DeMartini:02} which is in a good agreement with the
theoretical prediction $5/6 = 0.833$.

M.S. thanks Sean McConnell for his careful reading of the manuscript
and useful feedback. This work was supported by the DFG.


\begin{thebibliography}{18}
\bibitem{comment}
   Following the optimization procedure described in
   Section~\ref{sec:3}, one may find that the fidelity of the
   universal \textsl{C-NOT} gate for an arbitrary input states of
   qubits equals $F = \frac{5}{6} \times \frac{2}{3} = \frac{5}{9}$.
   This gate has structure similar to the universal quantum cloning
   machine \cite{Buzek:96}
\end{thebibliography}
\end{document}